\documentclass[preprint,journal]{VGTC/vgtc}       

\ifpdf
  \pdfoutput=1\relax                   
  \usepackage{graphicx}                
  \DeclareGraphicsExtensions{.pdf,.png,.jpg,.jpeg} 
\else
  \usepackage{graphicx}                
  \DeclareGraphicsExtensions{.eps}     
\fi%

\graphicspath{{figures/}{pictures/}{images/}{./}} 

\usepackage{microtype}                 
\PassOptionsToPackage{warn}{textcomp}  
\usepackage{textcomp}                  
\usepackage{mathptmx}                  
\usepackage{times}                     
\usepackage{cite}                      
\usepackage{tabu}                      
\usepackage{booktabs}                  
\usepackage{url}
\usepackage[dvipsnames]{xcolor}

\newcommand{\highlightChanges}[1]{\textcolor{black}{#1}}

\onlineid{1261}

\vgtccategory{Research}
\vgtcpapertype{system}

\title{P6: A Declarative Language for Integrating Machine Learning\\ in Visual Analytics \\
}

\author{Jianping Kelvin Li and Kwan-Liu Ma}
\authorfooter{
\item
 Jianping Kelvin Li is with University of California at Davis. E-mail: kelli@ucdavis.edu.
\item
 Kwan-Liu Ma is with University of California at Davis. E-mail: ma@cs.ucdavis.edu.
}

\shortauthortitle{Biv \MakeLowercase{\textit{et al.}}: P6}

\abstract{
We present P6, a declarative language for building high performance visual analytics systems through its support for specifying and integrating machine learning and interactive visualization methods. 
As data analysis methods based on machine learning and artificial intelligence continue to advance, a visual analytics solution can leverage these methods for better exploiting large and complex data. 
However, integrating machine learning methods with interactive visual analysis is challenging. 
Existing declarative programming libraries and toolkits for visualization lack support for coupling machine learning methods.
By providing a declarative language for visual analytics, P6 can empower more developers to create visual analytics applications that combine machine learning and visualization methods for data analysis and problem solving.
Through a variety of example applications, we demonstrate P6's capabilities and show the benefits of using declarative specifications to build visual analytics systems.
We also identify and discuss the research opportunities and challenges for declarative visual analytics.
} 


\vgtcinsertpkg

\begin{document}


\firstsection{Introduction}

\maketitle

Visual analytics systems combine interactive visualizations with data analysis and machine learning methods for empowering people to analyze, explore, and understand large data~\cite{cook2005illuminating}.
To build a visual analytics system, developers often need to use multiple programming libraries and toolkits to develop different visualization and analytics components, as well as to make them interoperable for supporting interactive visual analysis.
Programming with various libraries and toolkits often involves different computing languages and environments.
For instance, a web-based visual analytics application can employ JavaScript for interactive visualizations and Python for machine learning, which need to communicate in a server-client architecture or cloud computing infrastructure.
In general, building visual analytics systems requires developers to have not only knowledge in both visualization and machine learning but also skills for full-stack system development.

Most existing visualization toolkits~\cite{bostock2011d3, wickham2016ggplot2, satyanarayan2016reactive, satyanarayan2017vega} with declarative grammars can be used for expressive and rapid specifications of the visualization components.
However, these declarative toolkits do not support advanced analytics or machine learning methods.
Due to the lack of a common and unified method for integrating interactive visualizations with machine learning, a declarative programming approach for visual analytics is missing.
With high diversity and complexity in system development, building visual analytics systems is difficult and demands significant engineering efforts.
In addition, the increasing sizes and complexity in datasets impose more challenges to visual analytics system developments.
Many visual analytics applications need to process and visualize massive amounts of data, requiring both the visualization and analytics components to be high performance and scalable.
New toolkits are needed to support building visual analytics systems.

In our prior work, we developed P4 (Portable Parallel Processing Pipeline)~\cite{li2018p4}, a declarative visualization toolkit for building GPU-accelerated visualization systems.
We have also extended P4 to develop P5~\cite{li2019p5} for supporting parallel progressive visualization.
In this paper, we present P6, a programming toolkit using a declarative language to build visual analytics systems that can leverage the power of machine learning.
The design of P6 is motivated by three major design goals.\\

\noindent
\textbf{Interactive Machine Learning and Visualization.}
Machine learning methods are commonly used in visual analytics for facilitating automated analysis~\cite{sacha2016human, endert2017state}.
Interactive visualizations are used in complement to support human-computer interaction (HCI) and visual analysis.
P6 aims to provide support for implementing both machine learning and visualization methods, as well as integrating these methods in interactive visual interfaces.
The parameters for both machine learning and visualization methods should also be easily configured and refined.

\noindent
\textbf{Interactive and Scalable systems.}
Visual analytics demand high performance systems for processing and visualizing large datasets~\cite{wong2012top,steed2014web,sacha2016human}.
Visual analytics systems should leverage parallel processors and distributed computing to facilitate effective HCI and visual analysis processes.
To ensure interactive performance and scalability, applications created by P6 should be incorporated with high performance computing techniques for building efficient visual analytics systems.

\noindent
\textbf{Declarative Visual Analytics.}
Declarative grammars prove to be useful for creating interactive visualization applications~\cite{satyanarayan2016reactive, satyanarayan2017vega}.
Declarative visualization grammars can be extended to support designs and implementations of visual analytics systems.
Researchers and developers should be allowed to expressively and rapidly specify and refine design specifications as transiting from system design and implementation to deployment. 
We amid to develop a declarative programming toolkit for streamlining developments of visual analytics systems. 

By addressing the challenges to achieve these goals, we contribute 
a declarative language for rapidly specifying the design of visual analytics systems that integrate machine learning and visualization methods for interactive visual analysis.
For high performance, P6 leverages GPU computing to accelerate visualization renderings and allows machine learning computations to be offloaded to distributed computing systems.
We also contribute a framework that efficiently integrates machine learning methods with interactive visualizations. 
P6 makes it easy for researchers and developers to build visual analytics systems and applications.
The current implementation of P6 supports common machine learning methods, including clustering, dimension reduction,  regression, classification, and time-series analysis.
We demonstrate the usefulness of P6 with several case studies and show the benefits of using a declarative language for designing visual analytics applications.
\section{Related Work}
Our work is related to design models, system frameworks, and programming toolkits for visual analytics.

\subsection{Visual Analytics Systems}
Researchers have developed many models and frameworks for designing visual analytics systems.
HCI Models for sense making~\cite{pirolli2005sensemaking}, developing insights~\cite{keim2010visual}, and knowledge generation~\cite{sacha2014knowledge} are useful for high-level system design considerations.
Visualization models and frameworks ~\cite{shneiderman1996eyes,chi1998operator,munzner2009nested} are helpful for designing and crafting visual representations.
In addition, interaction models and taxonomies~\cite{yi2007toward,lam2008framework, pike2009science,heer2012interactive} provide guidance for supporting interactive visualizations and designing visual interfaces.
While these models and frameworks are certainly of great value for providing different levels of design considerations, effective toolkits for implementing visual analytics systems are lacking.
P6 is designed for allowing developers to easily follow the guidelines in these models and frameworks to build visual analytics solutions.
By using a well-designed declarative language, P6 provides a unified and expressive way to specify  components and operations in a visual analytics system for streamlining the process of system development. 

Machine learning algorithms can be leveraged for effective data analysis, and thus have increasing usages in visual analytics systems~\cite{sacha2016human,jiang2019recent}.
The most common machine learning algorithms used with visual analytics are: dimension reduction, clustering, classification, and regression~\cite{keim2015bridging,endert2017state}.
Visual analytics systems that employ these methods typically provide interactive visual interfaces for selecting algorithms, controlling the parameters of algorithms, defining the measures for the computations, or defining the analytical expectations~\cite{endert2017state}.
To support effective incorporation of machine learning into visual analytics systems, P6 provides many common machine learning algorithms and allows them to be easily integrated with visualization and interaction methods. 

Besides system functionalities, performance and scalability are also important to visual analytics~\cite{fekete2013software, liu2014effects}.
To support interactive performance and scalability, researchers have developed methods for exploiting GPU computing to accelerate data processing and visualization rendering~\cite{fekete2002interactive,liu2013immens,ren2017stardust}.
Some researchers also leveraged distributed computing to improve system scalability and interactivity~\cite{chan2008maintaining,jo2017swifttuna}.
For visual analytics applications with large datasets, high-performance computing techniques, such as GPU computing and distributed computing, are required for executing analytics and visualization algorithms~\cite{cook2005illuminating}.
However, leveraging both GPU computing and distributed computing for visual analytics systems is difficult.
Most systems that employ these techniques are typically implement for specific algorithms or analysis tasks using ad-hoc approaches.

\highlightChanges{
In our prior work, we developed P4 to allow declarative visualization grammar to be used with GPU computing via WebGL.
The design of P4 is based on the InfoVis Pipeline model~\cite{chi1998operator}, in which it performs data transformation, visual mapping, and view transformation to generate visualizations.
Based on the declarative specification, P4 automatically creates efficient WebGL programs at runtime to facilitate interactive visualizations.
To allow declarative specification of visual analytics processes, P6 uses P4 for GPU-accelerated visualization and extends the declarative grammar to support advanced data analytics with integration of machine learning methods. 
P6 also allows data analytics operations to be offloaded to a computing cluster for distributed computing.
}

\subsection{Programming Toolkits for Visual Analytics}
\highlightChanges{
Declarative visualization grammars based on Wilkinson's Grammar of Graphics~\cite{wilkinson1999computing} allow programmers to focus on the design of the visualization without worrying about the computation details of rendering~\cite{heer2010declarative}.
Popular visualization libraries and toolkits, such as ggplot2~\cite{wickham2016ggplot2} and D3~\cite{bostock2011d3}, typically provide declarative grammars for specifying visual encoding to create visualizations.
In addition to visual encoding, Vega~\cite{satyanarayan2016reactive} and Vega-Lite~\cite{satyanarayan2017vega} provide declarative interaction grammars for authoring interactive visualizations.
}
Since the web has become the most popular and convenient platform for deploying applications, most of these libraries and toolkits are based on JavaScript. 
These JavaScript libraries typically have functionalities for simple data transformations (e.g., filtering and aggregation), but they do not provide advanced statistical or machine learning methods for data analysis, nor provide support for integrating machine learning methods with interactive visualizations.

Researchers have also developed declarative languages for machine learning~\cite{ghoting2011systemml,boehm2016systemml}, which can be used with distributed computing systems, such as MapReduce~\cite{dean2004mapreduce} and Spark~\cite{zaharia2012resilient}, to execute machine learning algorithms.
However, these declarative languages are rather low-level, as their design goal is to express custom machine learning algorithms using statistical functions and linear algebra primitives.
Besides declarative languages, many programming libraries in Python provide off-the-shelf functions for applying machine learning algorithms to analyze data, such as Scikit-Learn~\cite{pedregosa2011scikit}, Statsmodel~\cite{seabold2010statsmodels}, and Spark MLlib~\cite{meng2016mllib}.
In addition, many numerical computing frameworks, such as Numpy~\cite{walt2011numpy} and Pandas~\cite{mckinney2012python}, can be used for preprocessing and wrangling data, which is useful for managing the input and output to/from Python based machine learning libraries.
These frameworks and libraries have made Python popular for machine learning and data science.
However, support for interactive visualization is lacking.
Due to the fact that most interactive visualization libraries are based on JavaScript and usable machine learning libraries are based on Python, the most conventional approach to build visual analytics systems is to use both JavaScript and Python.
This requires systems to incorporate inter-communication protocols between the two computing environments, such as HTTP or WebSocket over a server-client architecture.

Several Python libraries embed a JavaScript visualization library  to provide a wrapper API for creating interactive visualizations in Python without writing JavaScript codes.
These libraries also implement internal mechanisms for communicating between Python and JavaScript.
For example, Dash~\cite{dashvis} offers interactive chart templates in Python that use D3 for rendering the charts,
and Altair~\cite{vanderplas2018altair} leverages Vega-Lite to provide a Python API for declarative visualization specification.
These libraries allow Python programs to use JavaScript libraries for visualizations and machine learning libraries in Python for data analysis.
These libraries are also widely used in data science and computational notebooks~\cite{kluyver2016jupyter,radle2017codestrates,rule2018exploration}, which recently become a popular medium for data science.
However, this approach cannot be used for building web-based applications.
In addition, these libraries lack support for processing and visualizing large data, and their internal mechanisms for transferring data are difficult to be modified or replaced.

P6 uses a different approach as it is designed for building visual analytics systems that can be deployed on the web and other platforms.
P6 uses Python based libraries for supporting data analytics operations, which the functions for different machine learning algorithms can be used in P6's declarative language.
P6 provides a JavaScript API for web-based applications, and it supports a JSON file format for embedding in other languages or generating specifications from other systems.
Developers and programmers do not need to write Python codes when using P6 for building visual analytics systems, nor implementing the server-client components and communication mechanisms.
At runtime, the declarative specifications of machine learning and visualization methods are converted to programs for running on P6's system with an efficient mechanism to manage data transfers, inter-communications, and execution flow.
This mechanism also effectively supports the use GPU computing to accelerate visualization operations and offload machine learning computations to a computing cluster.

\section{Declarative Visual Analytics}\label{sec:design}
The core of P6 is a declarative language for specifying and integrating machine learning and interactive visualizations.
This declarative language allows P6 to incorporate a reactive workflow for implementing interactive visual interfaces.
In addition, P6 uses a unified data structure to effectively leverage GPU computing for accelerating visualizations and supports offloading machine learning computations to distributed systems.

\begin{figure}[t]
   \centering
   \includegraphics[width=\linewidth]{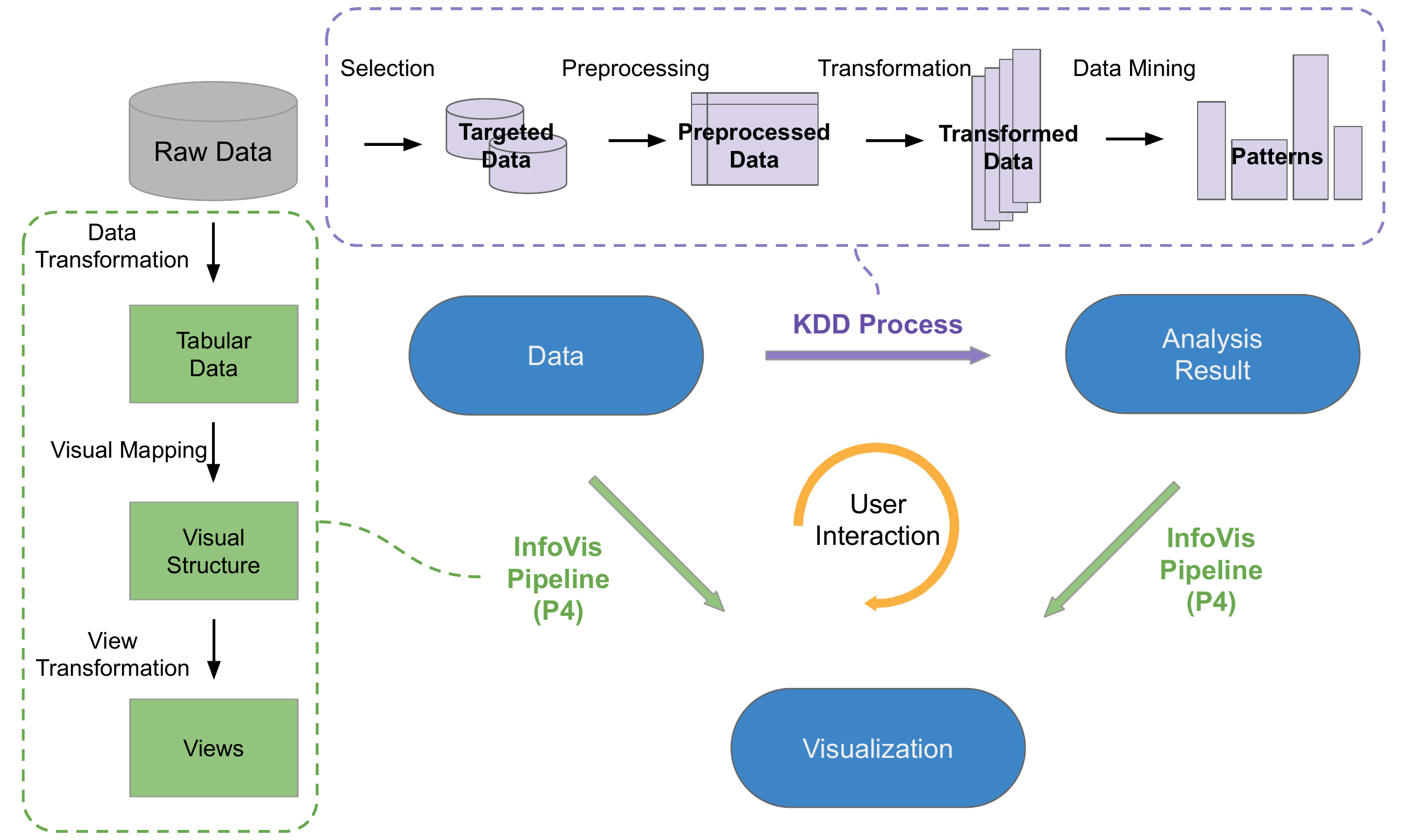}
   \caption{
   \highlightChanges{The design of the P6 declarative language is based on a conceptual framework that consists of the KDD Process model for data analysis and the InfoVis Pipeline model for visualization. }
   }
    \label{fig:VAModel}
\end{figure}


\begin{figure*}[t]
   \centering
   \includegraphics[width=\linewidth]{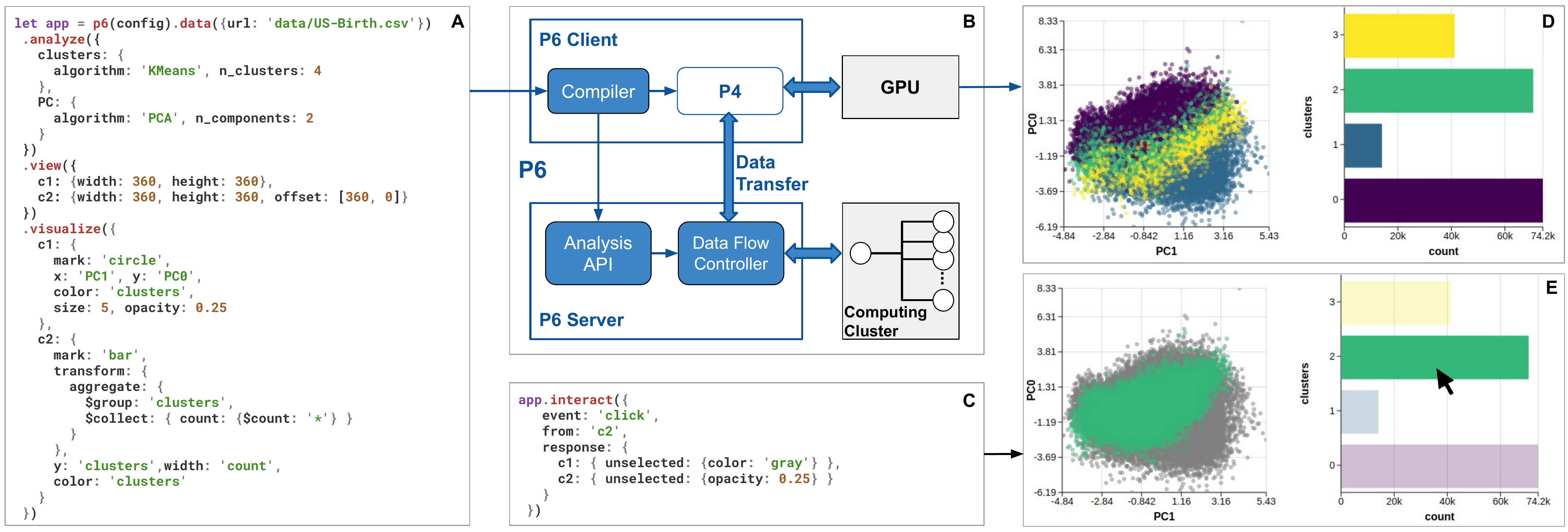}
   \caption{Declarative specification (A) of a visual analytics process that applies K-Means and PCA to analyze a dataset. The declarative specification is processed by P6's runtime (B). Interaction specification (C) can also be added to provide interactive visualizations for analyzing the K-Means and PCA results, as shown in the scatter plots and bar charts (D).}
    \label{fig:FrameworkDiagram}
\end{figure*}

\subsection{System Framework}\label{sec:model}
\highlightChanges{
The design of P6 is based on a conceptual framework (\autoref{fig:VAModel}) that is simplified from the knowledge generation model for visual analytics by Sacha et al.~\cite{sacha2014knowledge}.
In this conceptual framework, the Knowledge Discovery and Data Mining (KDD) Process model~\cite{fayyad1996data} is used as the reference for declarative specification of advanced data analysis and machine learning methods, which includes selection, preprocessing, and transformation of the input data.
For visualization, P6 uses P4 to perform the operations in the InfoVis Pipeline model~\cite{chi1998operator}, which can visualize both the data and analysis results.
For interactive visual analysis, user interactions can be added to control both the machine learning and visualization methods.
}

To support interactive performance and provide scalability for visual analytics, P6 employs distributed computing to parallelize data processing and machine learning methods.
\autoref{fig:FrameworkDiagram} shows P6's server-client architecture, in which the frontend client is for generating visualizations and the backend server is for processing data and analytics.
At runtime, the P6 compiler converts user-defined specifications to function calls that execute in both the client and server.
With P6, programmers do not need to worry about implementing visual analytics operations for server-client communications.
P6 automatically synthesizes requisite data flow and event logic across the client and server to effectively execute the operations in the declarative specifications.
For visualization operations, P6 leverages P4 to create efficient GPU programs based on the declarative specifications.
For the \textit{analyses} specifications, the operations are converted to API calls to the backend server and retrieve the results to the frontend. 
To apply machine learning algorithms on large datasets, P6 can parallelize and distribute computations on a computer cluster.

\subsection{A Declarative Language for Visual Analytics}
The P6 declarative language can be used to rapidly specify data analysis, visualization, and interaction methods.
Common machine learning methods are supported for data analysis, in which the analysis results can be easily used with interactive visualizations,
The basic unit of specification for P6 is a \textit{pipeline}.
Formally, a \textit{pipeline} is composed of a set of specifications:

\vspace{5pt}
\noindent
\textit{pipeline := \{data, analyses, view-layout, visualizations, interactions\}}
\vspace{5pt}

\noindent
The \textit{data} specification identifies a data source for input, which can be a CSV file in a local computer or remote server.
A set of \textit{analyses} can be specified to be performed on the input data.
The \textit{view-layout} specifies how to arrange the views that contain the \textit{visualizations}, which specifies the encoding for mapping the \textit{analyses} results and/or the input data to visual channels.
The \textit{interactions} specifies the interactions to be enabled on the visualized items, which can be zoom, pan, click, and hover.
In addition, \textit{annotations} can be specified to add markers, such as trend lines and labels, on the views.
P6 provides a JavaScript API for using its declarative language for building web-based visual analytics applications.
Specifications using the JSON format or file is also supported.

\highlightChanges{The specification of the input data to a \textit{pipeline} follows the KDD Process model which supports three data processing operations and has the following syntax: }

\vspace{5pt} 
\noindent
\textit{data := \{source, selection, preprocessing, transform\}}
\vspace{5pt}\vspace{5pt}

\noindent
\highlightChanges{
For a data \textit{source}, a \textit{selection} can be specified for selecting data columns by data attribute names or sampling data by specifying the number of data items or rows. 
The \textit{preprocessing} operation can be used for cleaning the data (e.g., dropping null values) and performing numerical or one-hot encoding for categorical data. 
The \textit{transform} operation can be used for aggregating, filtering, and sorting based on specific data fields.
}

The example in \autoref{fig:FrameworkDiagram}A shows a declarative \textit{pipeline} specification for analyzing a dataset containing information about newborn babies and their parents.
In this \textit{pipeline}, the input data is loaded from a CSV file, and two machine learning methods, K-Means Clustering and Principal Component Analysis (PCA), are used to analyze the data.
Two views are specified for containing the visualizations.
In the first view, the PCA result is shown in a scatter plot, where the color of the dots are based on the K-Means result.
In the second view, a data transformation is performed to group the data points based on the K-Means result and visualize a bar chart with each bar representing a group.
In addition, user interactions can be added to the visualization, as shown in \autoref{fig:FrameworkDiagram}C.

\highlightChanges{
With the P6 declarative language, the components of a visual analytics \textit{pipeline} can be specified in any order. Multiple analysis, visualization, and interaction methods can be iteratively added to the \textit{pipeline}.
P6 determines the execution order based on the dependencies in the operations.
This allows P6 to support the nonlinear and iterative development process of visual analytics applications.
}

\begin{figure}[t]
   \centering
   \includegraphics[width=\linewidth]{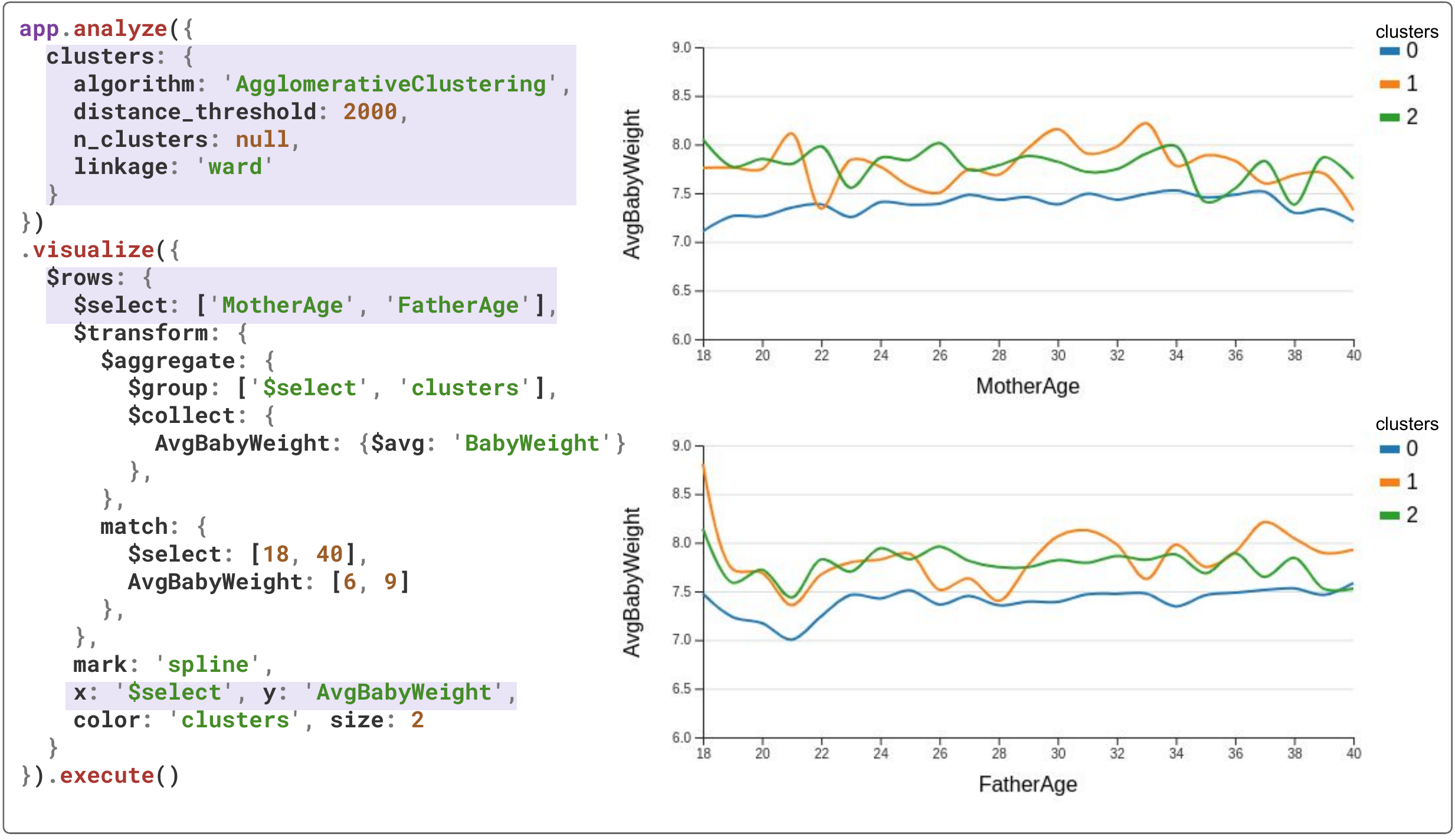}
   \caption{Declarative specification for using agglomerative clustering with data transformations and a multi-view layout. The clustering result is aggregated and filtered to show patterns and trends in a stack of two line charts.}
    \label{fig:HierClusteringExample}
\end{figure}

\begin{figure}[t]
   \centering
   \includegraphics[width=\linewidth]{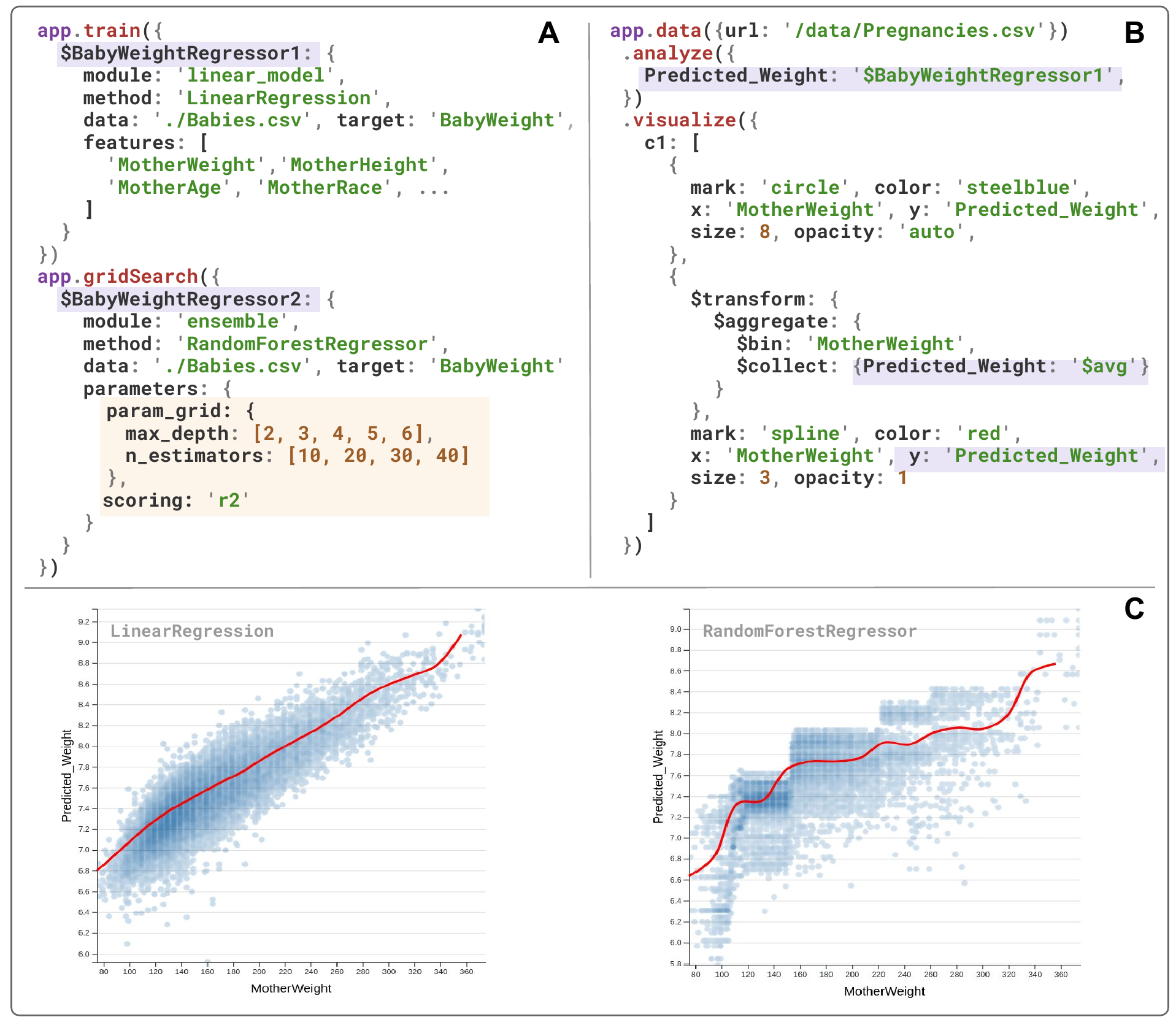}
   \caption{\highlightChanges{
   Declarative specification for training a linear regression model and a random forest model for predicting the baby weight based on the parent information (A). The trained models can be used for specifying analysis and visualization operations in P6 
   \textit{pipelines}(B). For visual analysis, additional aggregation can be performed on the prediction results to show trends and patterns (C). }
   }
    \label{fig:RegressionExample}
\end{figure}

\subsubsection{Data Analysis with Machine Learning}
P6 currently supports common machine learning methods for clustering, dimension reduction, regression, and time-series analysis.
An \textit{analysis} operation can be specified as follow:

\vspace{5pt} 
\noindent
\textit{analysis := \{features, scaling, algorithm, [parameters]\}}
\vspace{5pt}

\noindent
\highlightChanges{
The specification begins with defining a new name for the result to be computed using the chosen machine learning \textit{algorithm}.
The \textit{features} option is used for selecting which data attributes to use as the input to the machine learning algorithm.
The \textit{scaling} option is for scaling the input features using common methods such as normalization, scaling to unit variance, or scaling to a given range.
In addition, a set of parameters can be defined for the chosen \textit{algorithm}.
}
\autoref{fig:HierClusteringExample} shows another example, in which Agglomerative Clustering is used instead of K-Means.
With Agglomerative Clustering, the number of clusters does not need to be specified, because the algorithm determines the number of clusters based on the distance threshold parameter.
The analysis result of the machine learning technique can be easily used in visualizations.

In addition to the off-the-shelf machine learning algorithms, models for classification and regression can be trained and defined for predictive data analysis.
A classification or regression model can be defined as follows:

\vspace{5pt}
\noindent
\textit{model := \{module, method, training-data, target, features, [parameters]\}}
\vspace{5pt}
 
The \textit{module} and \textit{method} specify the system or library function for building the \textit{model}. The \textit{training-data} specifies data for training the model based on the \textit{target}, \textit{features}, and a set of \textit{parameters}.
\autoref{fig:RegressionExample} provides example specifications of two regression models based on the linear regression and random forest methods.
The models are trained to predict the baby birth weights (\textit{target}) based on the specified \textit{features}.
If no \textit{features} were provided, all the data dimensions and columns in the \textit{training-data} are used.
To use the trained model in analyses and visualizations, simply define a new output name in the \textit{analyses} specification, and point it to the name of the model.
This allows the model to be used for processing the input data and saving the result using the output name, as highlighted in purple color in \autoref{fig:RegressionExample}.
The result can be used for visualization by mapping it to a visual channel in the same way as using data attributes and analysis results.
The training of the model can be done offline, in which the visual analytics applications can use the trained model for predictions and analyses without requiring users to wait.

\begin{figure}[t]
   \centering
   \includegraphics[width=\linewidth]{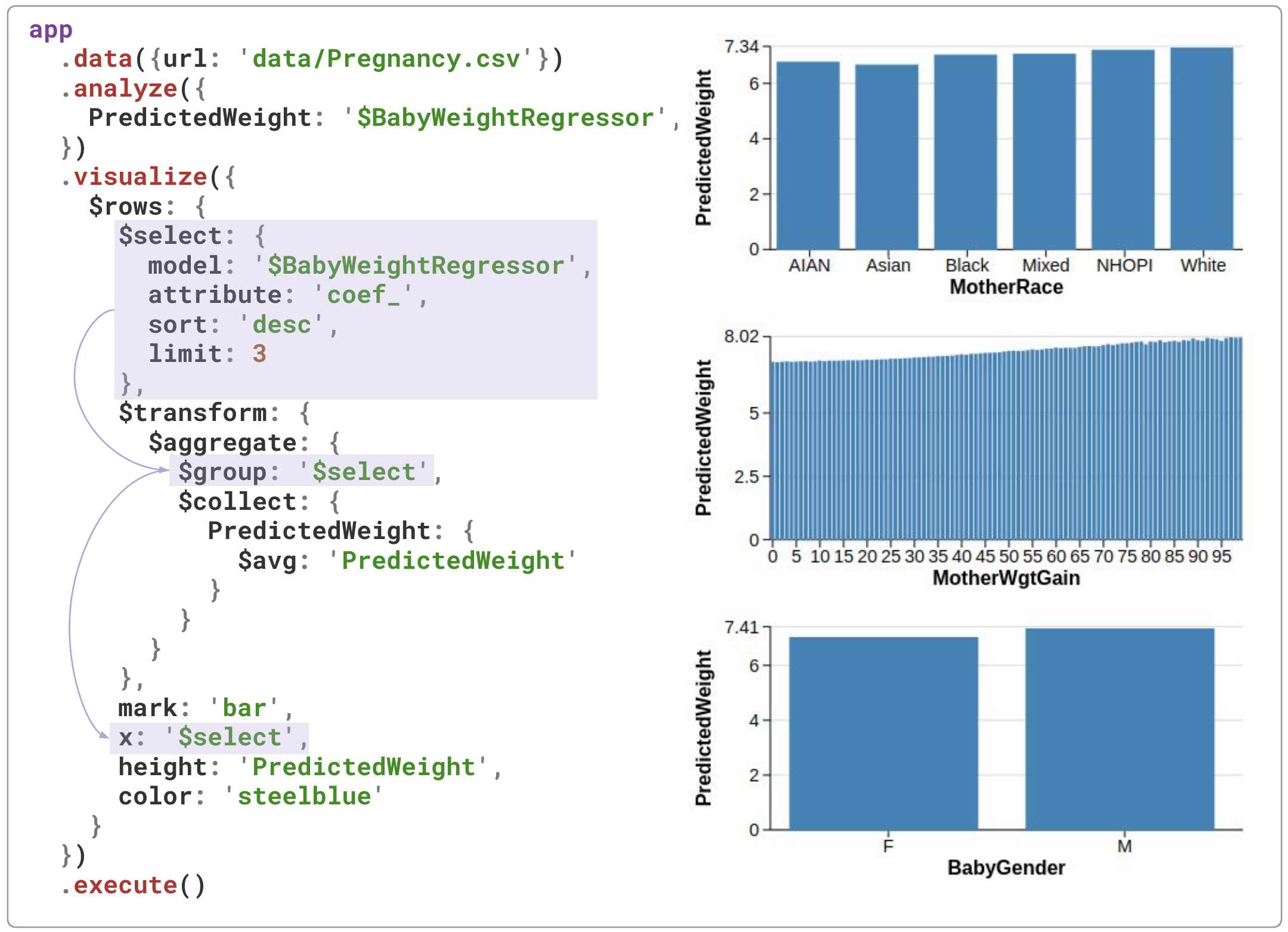}
   \caption{
   \highlightChanges{
   P6 allows using the attributes of the trained models for visualization. In this example, the coefficients of a linear regression model are used to visualize how the predictions vary with the top three most important features.}
   }
    \label{fig:TopFeatureExample}
\end{figure}

\begin{figure*}[t]
   \centering
   \includegraphics[width=\linewidth]{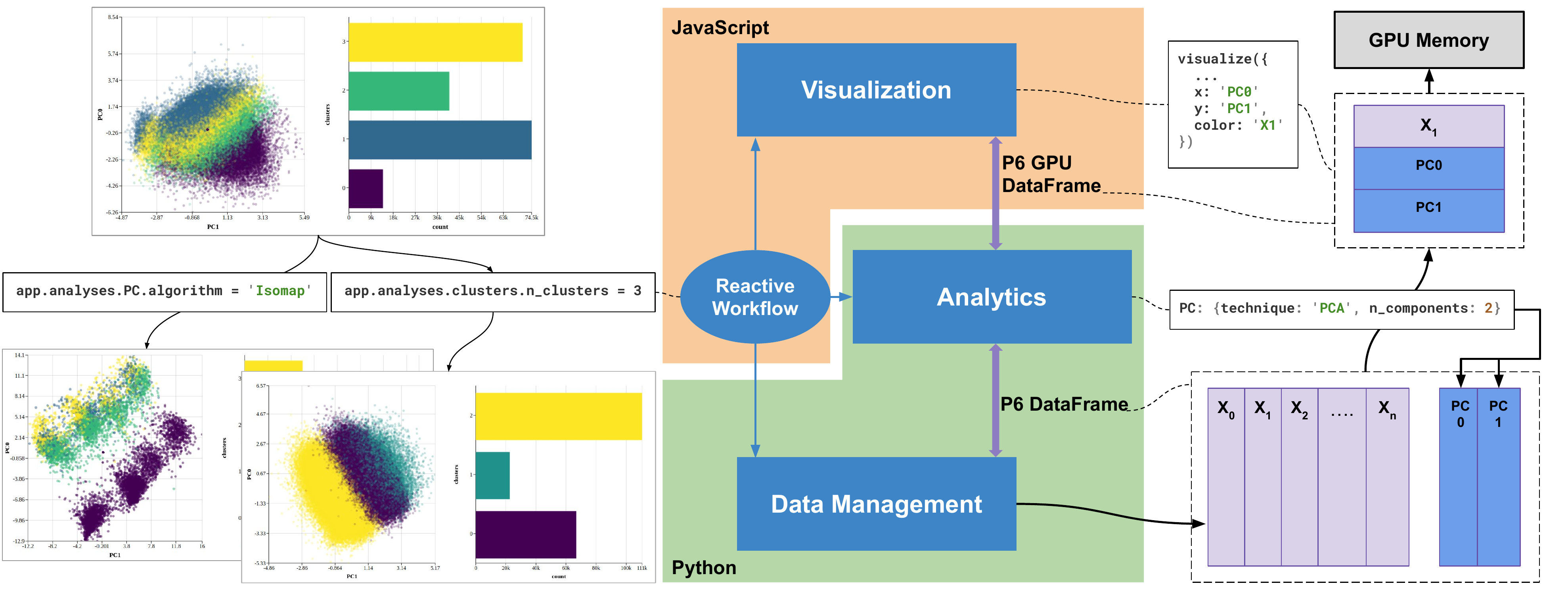}
   \caption{P6's reactive programming model for automatically updating all visualization views when the parameters were changed in the analysis and visualization methods. In P6's system architecture, a reactive workflow module is responsible to capturing the changes and notifies the data management, analytics, and visualization modules to execute the corresponding operations.
   A unified data structure is used for transferring data from the data management module to the analytics module, and sending the analysis result to the visualization module, where the result can be efficiently stored in GPU memory for processing and rendering.}
    \label{fig:ReactiveWorkflow}
\end{figure*}

\subsubsection{Integration of Machine Learning and Visualizations}
For visualization, the loaded datasets and the analysis results can be used together to generate multiple linked and interactive views using the layout defined in the \textit{view-layout} specification.
The specification of the visualization in each view can be defined as follows:

\vspace{5pt}
\noindent
\textit{view := \{transform, mark-type, [encodings]\}}
\vspace{5pt}

\noindent
Each visualization is specified by the view index.
A set of \textit{transforms} can be applied on the loaded data or results from the \textit{analyses}, including \textit{match} for filtering, \textit{derive} for calculating new attributes or metrics, and \textit{aggregate} for grouping or binning.
The \textit{mark-type} specifies the primitive geometric object for visually representing the data items.
The properties of these geometric objects are based on a set of \textit{encodings} that maps the attributes of the data items to visual channels, including position (x, y), size, width, height, color, and opacity.
\highlightChanges{
In addition to defining individual visualizations with view indexes, faceted views in rows or columns can be specified using the \textit{\$rows} and \textit{\$cols} operator, respectively.
The \textit{\$select} operator can be used to specify the variables for each of the views in rows or columns.
As highlighted in purple in \autoref{fig:HierClusteringExample}, the clustering result can be paired with different data attributes to explore trends and patterns in multiple views arranged in a column layout.
For appending multiple visualization items on the same view, the visualization specification for each view can be an array of specifications, instead of a single one, as the example shown in \autoref{fig:RegressionExample}.
}

\subsubsection{Tuning Parameters and Visualizing Model Attributes}

\highlightChanges{
For creating and training models for predictive analysis, P6 provides grid search to find the best parameter set for classification and regression models.
As shown in \autoref{fig:RegressionExample}A, grid search is used to find the best parameter set from the $param\_grid$ field using the r2 score metric.
Other metrics for scoring can be used, such as accuracy and F1 score.
In addition, P6 allows trained models to be saved as files.
At runtime, P6 \textit{pipelines} can load saved models from files for predictive analysis tasks. 
When using a trained model in visual analysis tasks, the attributes or metadata of the model can be used for creating visualizations to show the important aspects of a dataset.
Many regression and classification models provide information about the importance of each feature after training.
\autoref{fig:TopFeatureExample} shows an example of using the coefficients provided by the linear regression model for visualizing three bar charts which show how the predicted values vary across the top three most important features.
In this case, the top three features for the predicted baby weights based on the linear regression model are mother race, mother weight gained, and baby gender.
To create visualizations that show the predicted baby weights over these three features, the \textit{\$select} operator can be used to get the top three coefficients from the linear regression model, as shown in the  highlighted codes (purple) in \autoref{fig:TopFeatureExample}.
}

\subsubsection{Reactive Programming Model}\label{sec:reactive}
In visual analytics applications, users often need to modify or refine the parameters of the machine learning and visualization methods through a visual interface or by interacting with the visualizations.
For implementing such user interactions, reactive programming techniques can be adopted to automatically update the analysis and visualization results when any parameters were changed.
Reactive programming and workflows have been proven to be useful for information visualization systems~\cite{fekete2013software,fekete2013visual,satyanarayan2017vega}.
P6 combines reactive programming and declarative specifications for implementing user interface components in visual analytics applications.
For machine learning and visualization methods defined in the declarative specifications, properties and parameters of these methods can be modified in the program, which automatically trigger updates to the corresponding operations specified in the same \textit{pipeline}.
As the example shown in \autoref{fig:ReactiveWorkflow}, changes to the clustering methods or the method parameters automatically updates the associated visualizations with the clustering results.
This is particularly useful for implementing user interfaces to be used in visual analytics applications.
As visual analytics applications often need multiple data analysis and visualization methods, the declarative specification often has many operations defined.
Using the reactive programming capability in P6, the control logic associated with the user interfaces do not need to modify the reference of the declarative specification, as it can directly link to the parameter of the operations.
More details and examples are provided in our case studies.

\subsection{Implementation Details}
As P6 uses a server-client architecture, the backend server is implemented using Python, and the frontend client is implemented using JavaScript and WebGL.
The diagram in \autoref{fig:ReactiveWorkflow} shows the system model of P6.
The data management and the analytics module in the backend are built on top of the Python libraries, Pandas and Scikit-Learn~\cite{pedregosa2011scikit}.
Pandas is mainly used for loading, preprocessing, and managing data at the server side, and Scikit-Learn is used for the machine learning methods.
The names of the functions and parameters for the machine learning algorithms are the same in the Python libraries and declarative specifications.
For example, specifying the \textit{n\_clusters} for a clustering algorithm in the declarative specification directly passes the same parameter name to Scikit-Learn at the server-side.
At the frontend, P4 is used for GPU-accelerated visualizations, and we use the JavaScript Proxy API for implementing the reactive workflow module to track changes in the data, analytics, and visualization specifications, as well as to trigger the corresponding computations in both the server and client to update all the views.
To leverage distributed computing for accelerating machine learning algorithms, P6 uses Dask~\cite{dask}, a Python based distributed computing library, to allow the computations of Scikit-Learn functions to be offloaded to a specified computing cluster.

\subsubsection{Input and Output}\label{sec:dataio}
Visual analytics systems often need to load data from different sources and mediums.
P6 provides support for loading data in JSON or CSV formats from the server, client (e.g., web browser), and remote locations (e.g., external servers or web services).
For remote locations, P6 can fetch the data to the server. 
For loading data from the client side, P6 sends the data to the server side for executing the specified machine learning algorithms and passes the results back to client side for visualization.
For JavaScript applications, the analysis results can be exported as JSON for use in other web-based libraries. 
In addition, the \textit{exportAsJson} function can be used to export the result to the JSON format for each operation in the \textit{pipeline}.
Multiple P6 \textit{pipelines} with different data inputs can be specified, and the results of one \textit{pipeline} can be used in the other \textit{pipeline} for additional analyses and visualizations.

\subsubsection{Data Flow}
To ensure high performance, we have developed an effective data flow mechanism for managing and transferring data among the P6 client, server, and computing cluster.
To efficiently manage and transfer data, we extend the Pandas DataFrame to implement the P6 DataFrame as a unified data structure for data transfer. 
The P6 DataFrame can be transferred or streamed as binary data between the client and server without the need for data rearrangement (i.e., converting to JSON for transferring).
In addition, the P6 DataFrame can be effectively converted to the P6 GPU DataFrame for storing the analysis results in GPU memory, which allows us to use the same data model for using WebGL to perform data transformation and visualization operations.
The diagram in \autoref{fig:ReactiveWorkflow} (right) shows the data flow that uses the P6 DataFrame.
Based on the data definition in the declarative specification, the data management module loads the data from the input source to create an initial DataFrame and send it to the analytics module.
The analytics module performs the specified machine learning algorithms and appends the result as additional columns to the same DataFrame.
Based on the \textit{visualization} specification, the visualization module retrieves the required columns in the DataFrame from the analytics module to create a GPU DataFrame for GPU-based parallel visualizations.
With P6's system framework and data flow mechanism, we can effectively create visual analytics operations at runtime based on the declarative specifications.
The operations can be efficiently run on the backend server and computing clustering for machine learning computations as well as leverage WebGL for visualizing massive amounts of data items.

\subsection{Extensions}\label{sec:extension}
\begin{figure}[t]
   \centering
   \includegraphics[width=0.7\linewidth]{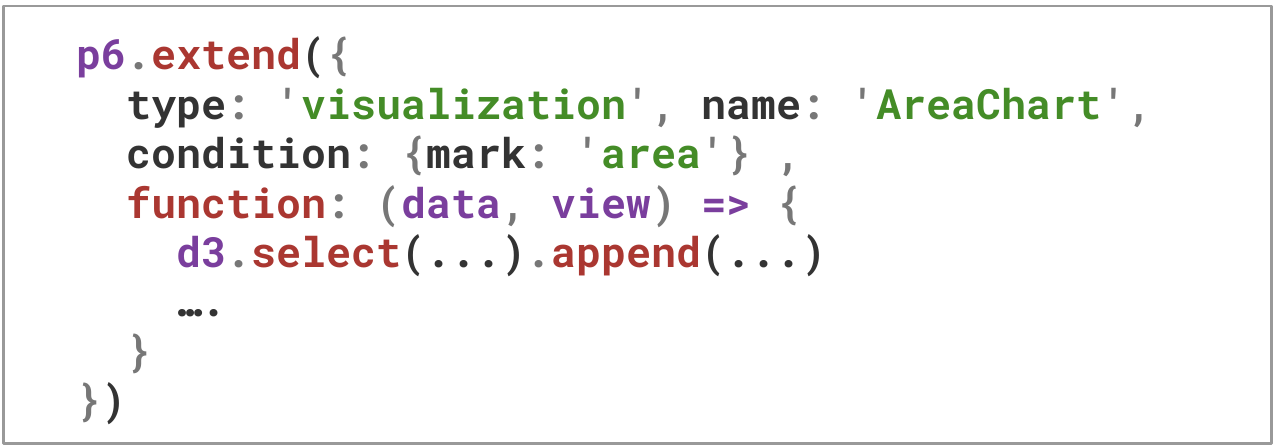}
   \caption{An example of using P6's visualization extension API to support area charts for using in declarative specifications. }
    \label{fig:VisExtensionAPI}
\end{figure}

P6 is designed to be flexible and extensible.
Both the analysis and visualization functionalities can be extended.
For visualization, custom chart templates can be added as plugins to P6 for creating custom plots.
P6 provides an extension API for specifying visualization plugins. 
The plugins can be used seamlessly with P6's declarative visualization grammar, including the GPU-based data transformations for processing the data before rendering the visualizations.
P6 can export the results of GPU-based transformations to JSON formats that can be used by other JavaScript visualization libraries.
\autoref{fig:VisExtensionAPI} provides an example of using the P6 extension API for adding a plugin function that uses D3 to implement an area chart.
The data and view setting are passed as inputs to the function.
The plugin can be used in P6's declarative visualization grammar by specifying the \textit{mark-type} to be 'area' as defined by the \textit{condition} property in P6's extension API.
To provide better functionalities for interactive visualizations beyond the capabilities in P4, we have added support for common charts used for information visualization.

For data analysis, P6 mainly uses Scikit-Learn to support machine learning methods.
Other data analytics and machine learning libraries that are based on Python and Numpy can also be added as plugins to P6's server backend.
However, distributed and parallel computing is not supported for executing the plugins for data analysis, but P6 can still effectively send the analysis results to the visualization frontend using P6 DataFrame, which can efficiently utilizes the GPU for data transformations and visualizations.

\section{Example Applications and Case Studies}\label{sec:applications}
In this section, we present example applications and case studies to demonstrate the capabilities and usefulness of P6 for developing visual analytics systems.

\begin{figure*}[t]
   \centering
   \includegraphics[width=\linewidth]{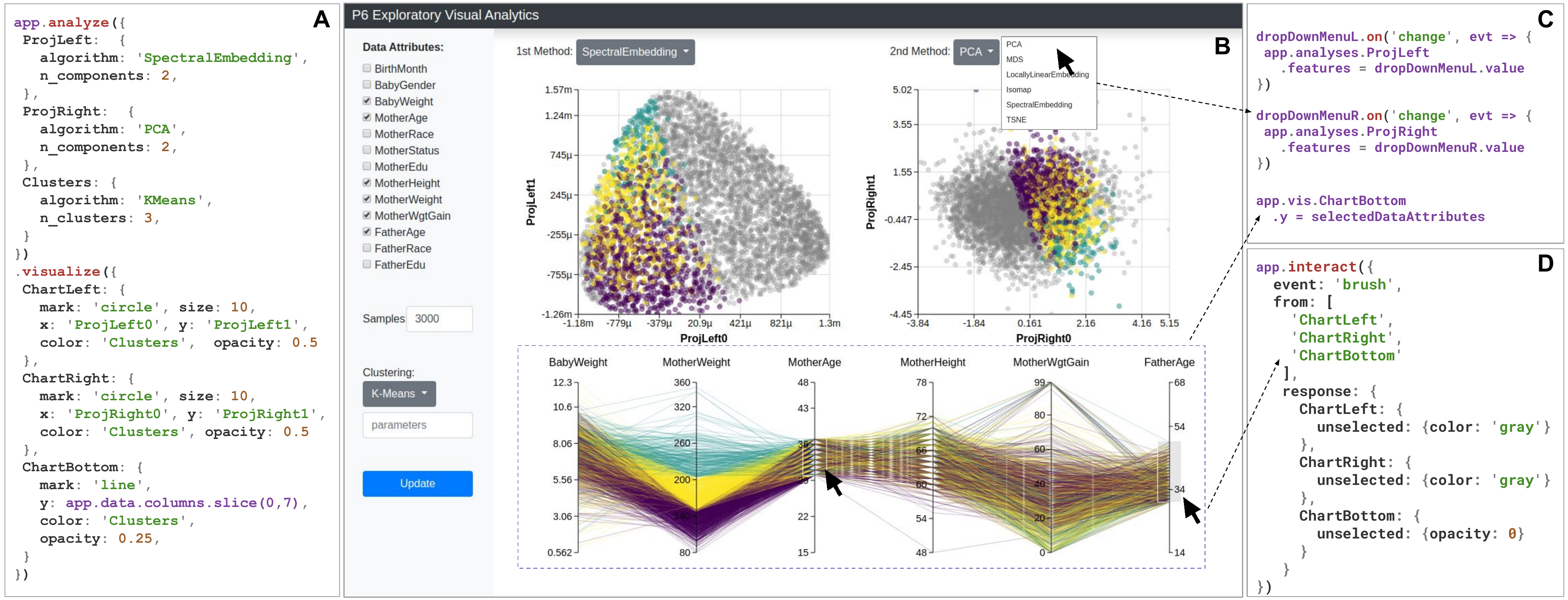}
   \caption{A visual analytics application for exploratory data analysis developed using P6's declarative language (A). The dashboard with the specified visualizations and visual interface is shown in (B). A reactive programming approach (C) is used to allow the visual interface to automatically update the analysis results and visualizations. The specification of interactions (D) allows users directly select data on the visualizations to better interpret and understand analysis results based on machine learning. }
    \label{fig:EvaApp}
\end{figure*}

\subsection{User Interface for Exploratory Visual Analysis}
In this case study, we evaluate P6's usefulness for creating visual interfaces for exploratory visual analysis (EVA).
An interactive visual interface is the core component of a visual analytics system.
To explore a dataset, analysts typically need to use multiple machine learning algorithms as well as compare different analysis methods.
We have developed a visual analytics application using P6 for allowing analysts to explore data using multiple dimension reduction methods with interactive visualizations.
\autoref{fig:EvaApp} shows the visual interface and dashboard of the application.
A set of initial analysis results and visualization views are displayed based on the specifications in (A).
Here we show the same dataset that we used in \autoref{sec:design}, but the application can be used with any multidimensional tabular datasets.
In the visual interface (B), analysts can select the data attributes and sampling size of the loaded dataset.
Optionally, a clustering method can be selected with custom parameters for analyzing the data along with the dimension reduction methods.
Two drop down menus are provided for selecting two different dimension reduction methods for projecting the dataset on the scatter plots.
The parallel coordinates plot at the bottom shows all the selected data attributes with each line representing a data item.
The specified reactive workflow (C) automatically updates the scatter plots when the selections in the drop down menus are changed.
Similarly, new selections of the data attributes immediately update the projection results and the parallel coordinates plot.
At last, the specification of the brushing-and-linking interaction (D) allows users to filter data items in the plots and see the correlations between the two projections.
The interaction on the parallel coordinates plot can be used for selecting the data items based on the original data attributes, allowing better interpretation of the dimension reduction results.
The data processing and rendering operations associated with the specified interactions are accelerated via GPU computing in P6.
This allows the system to support interactive visual analysis of large amounts of visualized data items.

This application demonstrates that P6 can make the implementation of visual analytics interfaces easy and fast.
The support for reactive workflow is particularly useful to implement UI widgets for changing the parameters of analysis and visualization methods.
Programmers can use only one line of code for modifying the analysis or visualization properties in the event handler or callback function for the UI widget.
While this application is designed for trying and comparing different dimension reduction methods for exploratory data analysis, we can easily modify it to support comparing models for classification or regression in a similar fashion.

\begin{figure}[t]
   \centering
   \includegraphics[width=\linewidth]{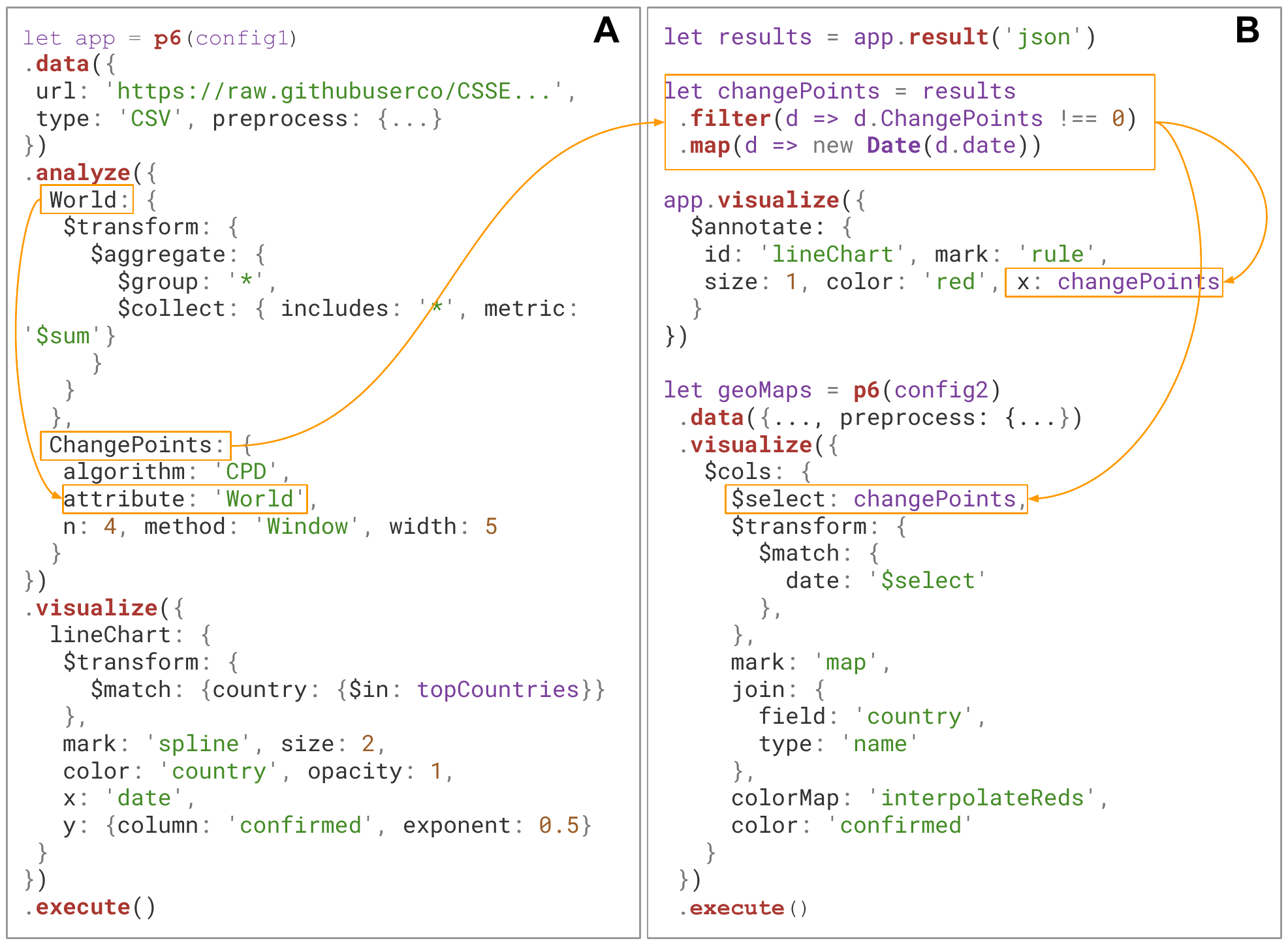}
   \caption{
   (A) Data aggregation and change point detection are used to analyze the time-series data and generate the timeline chart.
   (B) The result of a P6 \textit{pipeline} can be exported as JSON for annotation and used in another \textit{pipeline} for generated visualizations.
   }
    \label{fig:COVID19Codes}
\end{figure}

\begin{figure*}[t]
   \centering
   \includegraphics[width=\linewidth]{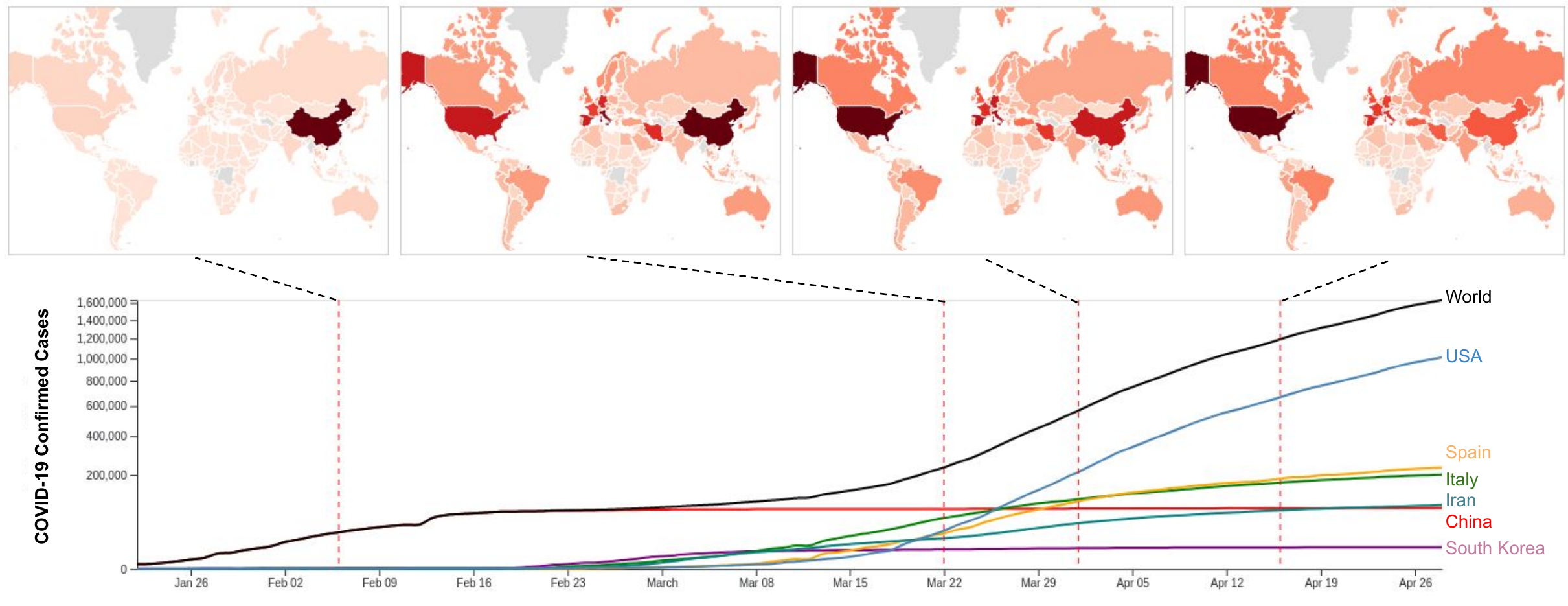}
   \caption{The timeline of the COVID-19 pandemic. Countries with high numbers of confirmed COVID-19 infections are selected to show in the line chart. The choropleth maps are generated based on the result of change point detection algorithm for providing snapshots about the spread of COVID-19 in different regions of the world.By using machine learning methods for change point detection and integrating with multi-view visualizations, P6 can be used to effectively generate explanatory visualizations to show the different stages of the COVID-19 pandemic.}
    \label{fig:COVID19UI}
\end{figure*}

\subsection{COVID-19 Global Cases}\label{sec:covid19}
Visual analytics can allow storytelling and explanatory analysis to better communicate information and knowledge.
In this case, we use the data source provided by Johns Hopkins University Center for Systems Science and Engineering for tracking the global numbers of confirmed  COVID-19 cases~\cite{COVID19DataGitHub}.
To analyze the spread of the virus over time, we use change point detection~\cite{truong2020selective} for finding the time points where the spread rate of COVID-19 have big changes.
The Python Ruptures library~\cite{truong2018ruptures} is used as a plugin in P6 for change point detection.

The declarative specification for using change point detection to analyze the COVID-19 dataset is shown in \autoref{fig:COVID19Codes}A.
As presented in \autoref{sec:dataio}, the data input for a \textit{pipeline} can be fetched from a remote server, and data preprocessing can be executed on the fetched data using P6's declarative grammars for data transformations.
To apply change point detection, we first aggregate the preprocessed time series to obtain the total global number of COVID-19 infections over time, and we use the aggregated time series as the input to the change point detection algorithm.
The change point detection result is exported as JSON and used to annotate the change points on the line chart and visualize the geographical heatmaps for each point.
\autoref{fig:COVID19Codes} shows the declarative specification for these operations.
As described in \autoref{sec:dataio}, the data analysis results can be exported as JSON and used as the input in another \textit{pipeline}.
To store the result in the P6 DataFrame, the change point detection algorithm marks a data item to be 1 if associated with a change point and otherwise 0.
Therefore, we need to filter the result based on the 'ChangePoints' column in the result for annotation on the line chart and generation of the choropleth maps.
The choropleth map visualizations use the \textit{\$cols} operator to specify the layout of multiple views in columns.
The declarative specifications of the two \textit{pipelines} in \autoref{fig:COVID19Codes} generates the visualizations in \autoref{fig:COVID19UI}.
From the line chart and the choropleth maps, we can see that China has about 30,000 COVID-19 cases on February 6, and the number of cases for all other countries is small.
On March 22, the numbers of COVID-19 cases have increased globally, where the countries in Europe and North America have the highest number of new cases.
On April 1, the number of new cases in China is small, and the numbers of cases in the United States, Spain, and Italy have surpassed China.
On April 16, the growth rate of COVID-19 cases has been slowed for countries in Europe, while the growth rate for the United States remains about the same. 

This application demonstrates that P6 can be useful for leveraging machine learning and visual analytics for storytelling and generating explanatory visualizations.

\subsection{Visual Analytics for High-Performance Computing}
\begin{figure}[t]
   \centering
   \includegraphics[width=\linewidth]{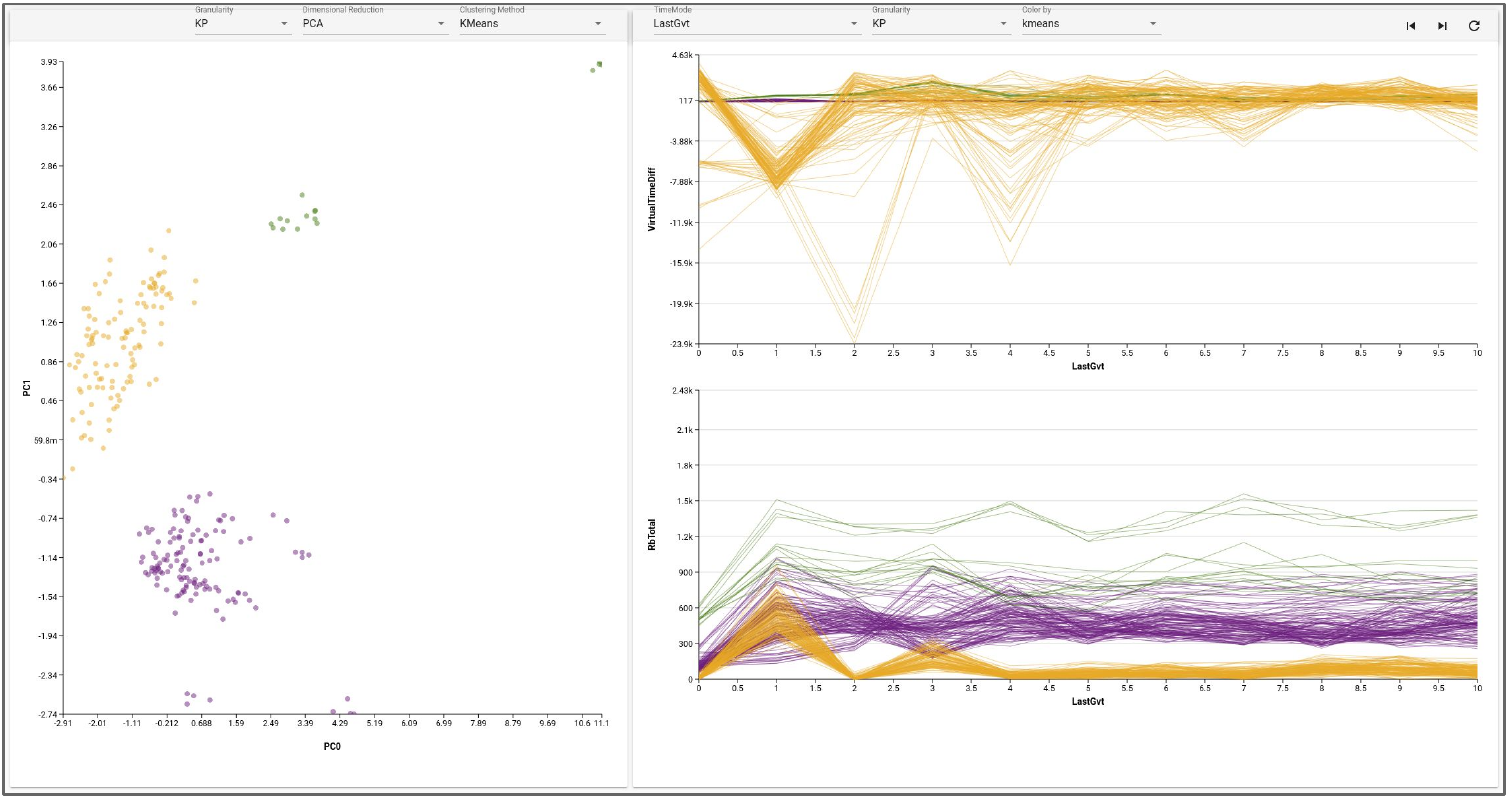}
   \caption{A visual analytics application developed by P6 for analyzing HPC time-series data. Clustering methods are applied to analyze the computing entities of an HPC application, in which the results are used for color encoding in the line charts to analyze the performance and temporal behaviors of each entity. }
    \label{fig:PdesApp}
\end{figure}

Developing visual analytics systems often requires domain-specific knowledge and collaboration with domain experts.
\highlightChanges{
We have used P6 to build a visual analytics system for assisting high-performance computing (HPC) researchers to analyze the performance and temporal behaviors of HPC applications.
}
HPC performance analysis typically requires exploration of multivariate time-series data for gaining insights to remove bottlenecks and improve application performance.
Various performance metrics (e.g. CPU and memory usage) at different levels of granularity (e.g., thread or compute node) are logged in the multivariate time-series data.
Prior works for analyzing HPC time-series~\cite{muelder2016visual, fujiwara2018visual} have shown that identifying the groups of compute nodes with different patterns and behaviors helps understanding and resolving performance issues.
By collaborating with HPC researchers, we designed visual analytics interfaces for supporting performance analysis of HPC applications.
\autoref{fig:PdesApp} shows a visual analytics interface for analyzing an HPC application running on a distributed system.
For parallel and distributed computing, this HPC application is executed with 16 processing entities (PE) running across the compute nodes, and each PE has 16 kernel processing (KP) running on different threads within a compute node.
Our visual interface allows selections of clustering methods for analyzing the application performance and behaviors at the KP or PE level.
The selected clustering method uses the average or accumulated values of each performance metric for each entity (KP or PE) to identify the groups with different behaviors.
To allow users to easily see and select the entities, PCA is used to project the entities on the scatter plot.
The clustering result is used for color encoding in the line charts.
Each line chart shows a performance metric for all the entities.
We can see that the HPC applications have three groups of entities with very different performance and temporal behaviors, and the line charts show the correlation between the two performance metrics. 
The declarative specification for this visual interface is similar to the examples in \autoref{fig:FrameworkDiagram} and \autoref{fig:HierClusteringExample}.
The extra specification needed is the data transformations for preprocessing the time-series data to obtain the aggregated values for each entity, in which the result is used as the input to the clustering algorithm.
To allow users to select different levels of granularity and clustering methods, we use the reactive programming capabilities in P6 for implementing the UI widgets in a similar approach described in \autoref{fig:ReactiveWorkflow} with specification similar to \autoref{fig:EvaApp}C.

Based on our experience, we found that collaborating with domain experts to develop visual analytics systems can be benefited from using a declarative language.
Domain-specific datasets typically need preprocessing before using as inputs to the analysis and visualization methods. 
By specifying the data preprocessing operations using declarative grammars, the domain experts can effectively review the preprocessing operations.
As the design and development of visual analytics systems is often an iterative process, declarative specifications can allow domain experts to understand what changes and improvements have been made at each iteration.
By trying the prototype system for providing feedback at each iteration, the domain experts can also slightly modify the declarative specifications to try different design variations or parameters for analysis and visualizations.
This allows the domain experts to provide more feedback that is useful for improving the system design.

\section{Discussion}\label{sec:discussion}
The examples and case studies demonstrate the capabilities and advantages of P6 for rapidly specifying and integrating machine learning and visualization methods.
Here, we discuss P6's current limitations and possible future work.

\subsection{Development and Debugging}
\highlightChanges{
With declarative specification for visual analytics, P6 can be beneficial to different types of application developers.
Application developers having a clear design specification can use the P6 declarative language to implement a visual analytics pipeline by completing one stage (\textit{data}, \textit{analyses}, \textit{view-layout}, \textit{visualizations}, or \textit{interactions}) at a time.
The declarative specification for each stage can be verbose for complex designs, so completing the \textit{pipeline} stage by stage is a better approach.
For visual analytics researchers to develop prototype systems, P6 can be used to first implement a simple \textit{pipeline} with all the stages specified.
The specifications for each stage can then be iteratively refined or extended as they explore and evaluate different design variations.
}

\highlightChanges{
A drawback of using declarative languages is that debugging becomes more difficult as internal structures and executions are not exposed to programmers.
In P6, visual analytics operations are executed across a server-client architecture.
To help programmers identify errors easier, P6 automatically collects errors in both the backend server and frontend client programs as any errors arise.
P6 also traces errors and indicates which stage and operation in the \textit{pipelines} are linked to the root causes.
In addition, the intermediate result for each operation in the \textit{pipeline} can be exported as JSON (described in \autoref{sec:dataio}) for inspection.
These functionalities can help the debugging processing, but debugging visual analytics applications with many interactions remains nontrivial.
While the benefits of declarative languages outweighted its drawback in debugging, it is worthwhile developing advanced tools to better support debugging and development. 
For example, Hoffswell et al.~\cite{hoffswell2016visual} have developed a visual debugging tool for interactive data visualization.
Their approach can be adopted and extended for debugging declarative specifications of visual analytics pipelines.
}

\subsection{Complex Visual Analytics Processes}
\highlightChanges{
P6's declarative language is designed to be simple and intuitive for implementing visual analytics applications.
As P6 supports many common machine learning and visualization methods, P6 is particularly useful for building interactive systems for common visual analysis tasks.
For data analysis or visualization methods that are not supported in P6, P6's plugin and extension interface described in \autoref{sec:extension} can be used to add custom operations.
In addition, multiple P6 \textit{pipelines} can be used together to support more complex visual analysis processes, as demonstrated in \autoref{sec:covid19}.
}

\highlightChanges{
For interactive visualizations, P6 currently supports common interaction methods, such as click, hover, brush, zoom, and pan.
In the P6 declarative language, the interaction specification can only filter data at the visualization level.
As to declarative interaction design for modifying data inputs and parameters of the data analysis methods and machine learning models, the support in P6 is limited.
For such analysis-level interactions, the reactive programming functionality described in \autoref{sec:reactive} can be leveraged as a workaround.
However, a declarative programming approach is desired.
In future work, we can extend the P6 declarative language for specifying both visualization-level and analysis-level interactions as well as distinguishing between them.
}


\highlightChanges{Our work is only an initial step toward declarative visual analytics.
P6's simplicity and advantage for implementing visual analytics applications should allow easy adoption by application developers and visual analytics practitioners.
In future work, we plan to collect user feedback to evaluate our design and improve our toolkit.
}


\subsection{Functionalities and Scalability}
Our approach of using a declarative language for integrating interactive visualization and machine learning can be applied to different architectures.
Although the use of P4 and a Python backend provides high performance for visualization and high functionalities for machine learning, systems can use different combinations of libraries and toolkits.
For example, we can replace the Python backend and Dask with Spark MLlib, which might provide better performance and scalability for machine learning computations.
Using declarative specifications without worrying about how to execute program operations with a server-client or distributed computing architecture allows more people to develop usable visual analytics systems.
However, the performance and scalability depends on the machine learning algorithms, so the system response time can have high variation for different algorithms.

Many machine learning algorithms have high computational complexity, which become slow when executing on large datasets.
Progressive visual analytics provides an effective way for interactive analysis of big data.
By delivering incremental results to maintain an acceptable level of responsiveness, progressive visual analytics systems allow users to interact with the intermediate results and steer the analysis process.
A challenging and promising direction for future research is to integrate machine learning with progressive visual analytics for interactive analysis of big data.
While not all machine learning methods can generate meaningful incremental or partial results, researchers in visual analytics and machine learning should collaborate to adapt more methods to support progressive data analysis.
For machine learning methods that can produce meaningful incremental results, declarative grammars can be designed for specifying how to use these methods for progressive and interactive visual analysis.
In P5~\cite{li2019p5}, we have developed a declarative language for progressive visualization, but it lacks support for using machine learning methods for advanced data analysis.
A declarative language for integrating incremental machine learning methods with progressive visual analytics can provide great value to researchers and developers.
System development can be benefited by providing rapid specifications for how to execute machine learning methods progressively, visualize the incremental results, adjust parameters in both the machine learning methods, and steer the visual analysis process.

\subsection{Toward Visual Analytics as Services}
When using the P6 declarative language for building a visual analytics system, 
programmers do not need to deal with the complexity of setting up the backend server or remote computing services.
This approach can be extended to support visual analytics as services that leverage cloud computing for performing data processing and machine learning computations.  
The operations in the declarative specifications are executed at runtime using cloud computing services.
Cloud computing can dynamically allocate computing instances based on the declarative specifications, providing  better scalability and availability for hosting visual analytics applications. 
We can extend the declarative language for requesting cloud computing resources based on the data size and computational complexity.
The request can be specified using the response time requirements for visual analysis tasks based on appropriate human computer interaction models~\cite{card1983psychology,card1999readings} and recent user study~\cite{liu2014effects}.
Providing visual analytics as services can make building scalable systems easier and allow better utilization of computing resources.



\section{conclusion}

We introduce P6, a toolkit providing a declarative language for building visual analytics systems that richly integrate machine learning and interactive visualizations.
The examples given in this paper have demonstrated the advantage of P6 for creating visual analytics applications.
With declarative specification for visual analytics, it is possible for non-experts 
to develop 
advanced data analysis and communication solutions that combine the strengths of human and artificial intelligence. 
We have also identified and discussed future research opportunities based on our work for improving and extending declarative visual analytics to contribute more successful designs of visual analytics systems.
The source code, documentations, and examples of P6 can be found at \url{https://github.com/jpkli/p6}.

\acknowledgments{
This research is sponsored in part by the U.S. National Science Foundation through grant 
IIS-1741536 and IIS-1528203, and also by the U.S. Department of Energy through grant DE-SC0014917.
}

\bibliographystyle{VGTC/abbrv-doi}

\bibliography{references,vast,ML,cova}
\end{document}